\documentstyle [epsf,spie,twocolumn] {article}
\input epsf

\title {Measurements with the Chandra X-Ray Observatory's flight contamination monitor\\ }

\author {R. F. Elsner, J. J. Kolodziejczak, S. L. O'Dell, D. A. Swartz, A. F. Tennant and M. C. Weisskopf
\skiplinehalf 
NASA Marshall Space Flight Center, SD50   \skipline 
Huntsville, AL \hspace{0.5em}35812 \hspace{0.5em}USA 
\skiplinehalf 
}

\authorinfo{{in {\em X-Ray Optics, Instruments, and Missions},  J.Truemper and B. Aschenbach, eds., {\em Proc.\ SPIE}\/ {\bf 4012}, 2000.}\\
R.F.E.: elsner@avalon.msfc.nasa.gov; 256-544-7741. \\
S.L.O.: odell@cosmos.msfc.nasa.gov; 256-544-7708. }

\newcommand{\lae}{\mathrel{<\kern-1.0em\lower0.9ex\hbox{$\sim$}}}
\newcommand{\gae}{\mathrel{>\kern-1.0em\lower0.9ex\hbox{$\sim$}}}

\begin{document}
\maketitle

\begin {abstract}
NASA's Chandra X-ray Observatory includes a Flight Contamination
Monitor (FCM), a system of 16 radioactive calibration sources mounted to the
inside of the Observatory's forward contamination cover. The purpose of the
FCM is to verify the ground-to-orbit transfer of the Chandra flux scale,
through comparison of data acquired during the ground calibration with those
obtained in orbit, immediately prior to opening the Observatory's sun-shade
door. Here we report results of these measurements, which place limits on
the change in mirror--detector system response and, hence, on any
accumulation of molecular contamination on the mirrors' iridium-coated
surfaces.


\end {abstract}

\section {Introduction} \label{s:introduction}

\begin {figure} [htb]
\centerline {\epsfxsize=0.95\hsize \epsffile{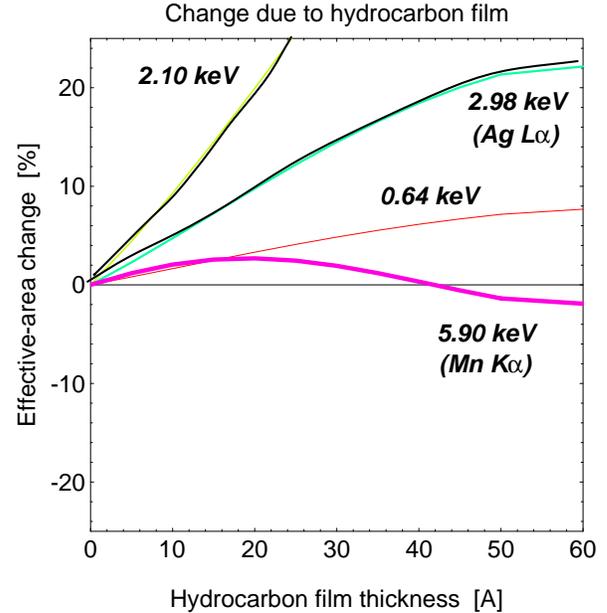} }
\caption {\label{fig:sensitivity} 
Chandra effective area changes in percent, at selected x-ray energies, as a function of the thickness of a uniform hydrocarbon film over 95\% bulk density iridium.}
\end {figure}

The Chandra X-ray Observatory (CXO), the x-ray component of NASA's Great Observatories launched in July 1999~\cite{Weisskopf2000}, underwent the most extensive, by far, system-level calibration program in the history of high-energy astrophysics at the Marshall Space Flight Center (MSFC) X-Ray Calibration Facility (XRCF) during the winter and spring of 1997~\cite{Weisskopf1997}.  
One of the key goals was to determine the absolute flux scale to an accuracy of a few percent or less, and detailed analysis of effective area measurements has achieved this goal for the ground calibration~\cite{Schwartz2000}.
In addition, a high-fidelity ray trace program, maintained by the Chandra Mission Support Team and designed to model the performance of the High Resolution Mirror Assembly (HRMA) in great detail, uses the ground calibration results to predict the on-orbit performance.  
However, it is still important to verify the transfer of the ground-level calibration to on-orbit operation, particularly in view of the absence of any well characterized astronomical x-ray standard candles and the sensitivity of the Chandra effective area to changes in molecular and particulate contamination~\cite{Elsner1994} (Fig.~\ref{fig:sensitivity}).  

\begin {figure} [htb]
\centerline {\epsfxsize=0.95\hsize \epsffile{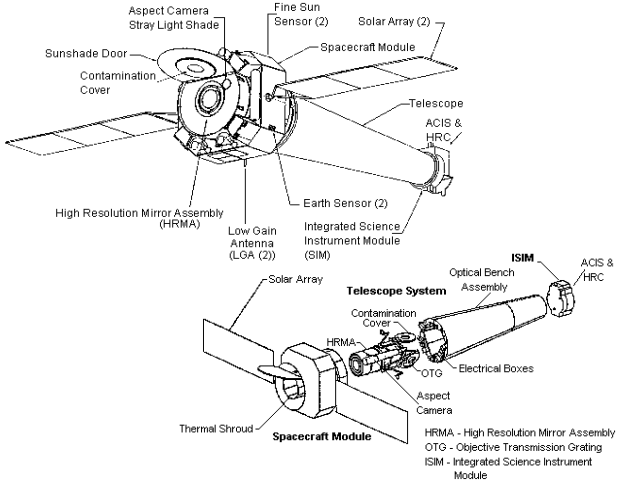} }
\caption {\label{fig:axaf} 
The Chandra spacecraft and components showing the FCC on the front of the HRMA.  FCM measurements are taken during the activation phase with the FCC closed.}
\end {figure}

For these reasons, a Flight Contamination Monitor system (FCM), made up of electron-capture radioactive sources emitting x-ray lines, is installed in the Forward Contamination Cover (FCC) of the HRMA (Fig.~\ref{fig:axaf}).
There are four sources, spaced 90$^o$ apart, for each HRMA shell, with activities and radii scaled appropriately for that shell's aperture.
The two objectives of the ground and in-orbit measurements using the FCM and focal plane detectors are:  

\begin {itemize}
\item [(1)] Verify the transfer of the HRMA absolute flux scale from the XRCF test phase to the orbital activation phase (OAC).
\item [(2)] Measure or bound any changes in molecular contamination of the HRMA.
\end {itemize}

\begin {figure} [htb]
\centerline {\epsfxsize=0.95\hsize \epsffile{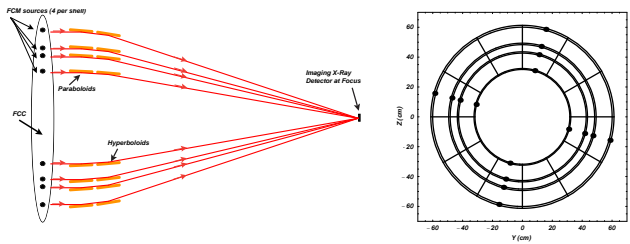} }
\caption {\label{fig:schematic} 
Left:  The x-ray optical path for FCM measurements showing the FCM sources in the FCC, reflections of the paraboloids and hyperboloids of the four HRMA mirror shells, and collection by a focal plane detector (ACIS and, on the ground only, ACIS-2C and HRC-I.  Right:  Positions of the FCM sources projected on the paraboloid apertures midway between support struts for the baffle plates in the thermal pre-collimator, central aperture plate, and thermal post-collimator.}
\end {figure}

The focal plane detectors, the Advanced CCD Imaging Spectrometer (ACIS) and the High Resolution Camera (HRC), also employ their own radioactive calibration sources, in order to monitor any change in detector performance.
During ground calibration FCM measurements were taken with ACIS, ACIS-2C (an ACIS surrogate), and HRC-I (the imaging readout for HRC).  The FCM measurements during the activation phase were the first look at the on-orbit HRMA/ACIS performance.  Figure~\ref{fig:schematic} is a schematic of the x-ray optical path for FCM measurements.

\begin {figure} [htb]
\centerline {\epsfxsize=0.95\hsize \epsffile{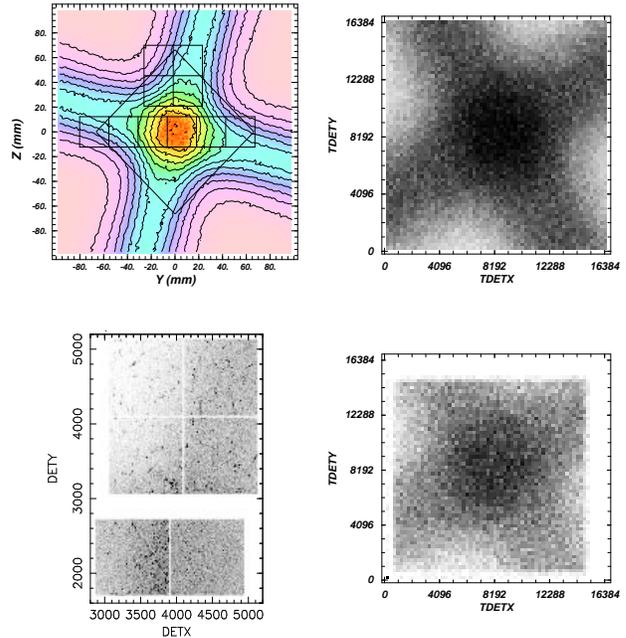} }
\caption {\label{fig:images} 
Simulated (top) and XRCF (bottom) FCM images.  The simulated image in the top left shows the cross pattern generated in the focal plane by the FCM system, with outlines of the focal plane detectors ACIS and HRC-I superimposed.  The image in the bottom left is a short ACIS exposure from XRCF.  On the right are HRC-I images in HRC detector coordinates.  On top is a simulation and on the bottom is an exposure from XRCF.}
\end {figure}

Each FCM source illuminates the portion of the paraboloid aperture immediately in front of it, leading to an image of a slightly curved stripe in the focal plane.  
The FCM source 180$^o$ away contributes a stripe superimposed on the first one, but with a slight curvature in the opposite direction.  
The pair of FCM sources at 90$^o$ with respect to the first two contribute stripes perpendicular to the first two, leading to an image of a cross in the focal plane (Fig.~\ref{fig:images}).  
Although the global topology of the image is independent of energy, intensity contours depend slightly on energy due to the differing energy response of the four HRMA shells.  Cross-correlation of measured and simulated images provides  a measure of any FCC position shifts (\S\ref{s:regis}).

Ideally, direct comparison of ground and on-orbit FCM measurements at as many lines as possible would provide a direct transfer of the absolute flux scale calibration from XRCF to orbit, or would provide direct measures of any discrepancies.
In practise, some modelling is necessary due to the differing HRMA/ACIS orientation at XRCF and OAC, the different gravity and thermal environments at XRCF and on-orbit, radioactive decay and ACIS dead time corrections, and possible changes in relative FCM, FCC and HRMA positions, for example due to launch vibrations.
Finally, interpretation of any discrepancies with the HRMA model derived from the ground calibration requires ray trace simulations, for which we use the Project Science ray trace code.

In this paper, we describe the characterization of the flight FCM radioactive sources (\S\ref{s:sources}), the spectral analysis of FCM measurements taken at XRCF and during OAC (\S\ref{s:spec}), the registration analysis necessary to determine any positional shifts of the FCC relative to its nominal position (\S\ref{s:regis}), and a comparison of the XRCF and OAC measurements with each other and with ray trace simulations, leading to the conclusion that any difference in HRMA throughput between XRCF and OAC is less than a few percent (\S\ref{s:constraints}).


\section{Radioactive sources} \label{s:sources}

\begin {figure} [htb]
\centerline {\epsfxsize=0.95\hsize \epsffile{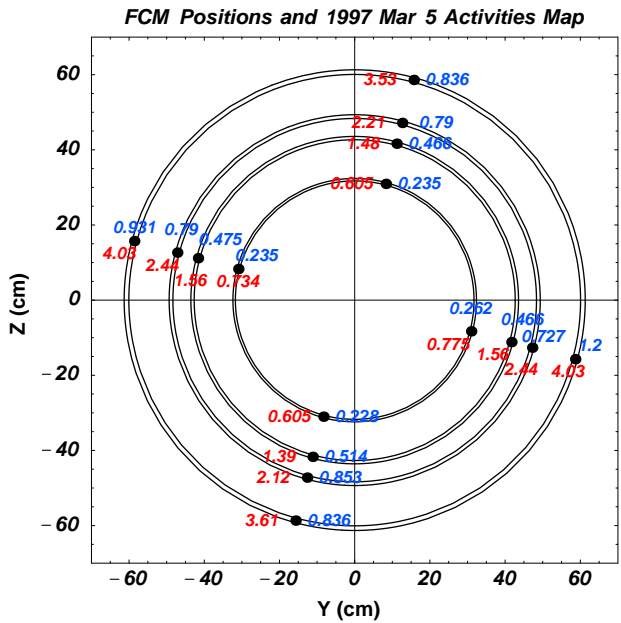} }
\caption {\label{fig:activities} 
Positions of the FCM sources projected on the paraboloid apertures.}
\end {figure}

The FCM comprises sixteen (one per shell per quadrant) electron-capture sources positioned midway between support struts (Fig.~\ref{fig:schematic}), containing ${}^{109}$Cd and ${}^{55}$Fe prepared at Isotope Products, Inc., packaged by MSFC, and mounted on the FCC.  
To prevent leakage, MSFC sealed each source assembly with a 100-$\mu$m-thick Be window, eliminating the detectability and utility of the Mn L$\alpha$ line at 0.64 keV.  
MSFC measured the individual FCM flight source activities (Table~\ref{tab:fcm} and Fig.~\ref{fig:activities}), with corroborating measurements by the US Army Redstone Arsenal.  
Dominated by systematic effects, the estimated errors in the measured absolute activities are $\sim$15\%.  
However, activity ratios between sources are known with significantly higher precision than the absolute activities.  
Statistical uncertainties (1$\sigma$) in the ${}^{55}$Fe activities, derived from measurements of the Mn K$\alpha$ line strength, range from 0.2\% (shell 1) to 0.4\% (shell 6).
Statistical uncertainties (1$\sigma$) in the ${}^{109}$Cd activities, derived from measurements of the Ag K$\alpha$ (22 keV) line strength, range from 0.04\% (shell 1) to 0.1\% (shell 6).
However, the lower energy ($\sim$3 keV) lines in the Ag L series are significantly affected by the overlying material in the source assemblies. 
The internal configuration, from the substrate up, is ${}^{109}$Cd ($\sim$0.3 $\mu$m), a gold buffer layer ($\sim$0.05 $\mu$m), ${}^{55}$Fe ($\sim$0.1 $\mu$m), a gold sealing layer ($\sim$0.05 $\mu$m), and finally the beryllium window (100 $\mu$m). 
These layers absorb $\sim$60\% of the 3 keV Ag L series x-rays emitted from the ${}^{109}$Cd.
The range over which the actual layer thicknesses vary about their nominal values is unknown.

Ray trace simulations for the registration analysis (\S~\ref{s:regis} below) use the activities given in Table~\ref{tab:fcm}.  
The radioactive half-lives of ${}^{109}$Cd and ${}^{55}$Fe are 1.2665+/-0.0011 and 2.73+/-0.03 yr~\cite{Browne1986}, respectively, so compensation for radioactive decay is necessary in the data analysis.  MSFC also mapped source uniformities (see ~\cite{Elsner1998} for examples of uniformity maps).  
These uniformity maps are now included in the Chandra Project Science ray trace code.

\begin {table} [htb]
\caption {Nominal FCM source source activities (mCi) on 1997 Mar 5${}^*$ and radii (mm).}
\label{tab:fcm}
\begin {center}
\begin {tabular}{|c|c|c|c|c|} \hline
{\bf } & 1 & 3 & 4 & 6 \\ 
\hline
\rule[-1ex]{0pt}{3.5ex}  -15  & 4.03,1.20 & 2.44,0.73 & 1.56,0.47 & 0.78,0.26 \\
\hline
\rule[-1ex]{0pt}{3.5ex}  75  & 3.53,0.84 & 2.21,0.79 & 1.49,0.47 & 0.61,0.24 \\
\hline
\rule[-1ex]{0pt}{3.5ex}  165  & 4.03,0.93 & 2.44,0.79 & 1.56,0.48 & 0.73,0.24 \\
\hline
\rule[-1ex]{0pt}{3.5ex}  255  & 3.61,0.84 & 2.12,0.85 & 1.40,0.51 & 0.61,0.23 \\
\hline
\rule[-1ex]{0pt}{3.5ex}  Avg.  & 3.8,0.95  & 2.3,0.79  & 1.5,0.48  & 0.68,0.24 \\
\hline
\rule[-1ex]{0pt}{3.5ex}  R  & 4.500 & 3.625 & 3.200 & 2.380 \\
\hline
\end {tabular}
\end {center}
\begin {itemize}
\item [${}^*$] The first number is for ${}^{109}$Cd and the second for ${}^{55}$Fe.
\end {itemize}
\end {table}

MSFC also used Monte-Carlo simulation of the source assembly to determine fluxes in the forward direction.
Predicted absolute fluxes for electron-capture lines are no more accurate than the absolute activity measurements, but relative fluxes should be as accurate as the tabulated branching ratios - a few \%~\cite{Browne1986}.
Predicted fluxes, absolute or relative, for fluorescent lines produced in the assemblies (primarily by the Ag K$\alpha$ line at 22 keV) are much less secure due to the complexity of the simulations and the number of interaction pathways.


\section {Spectral analysis} \label{s:spec}

\begin {figure} [htb]
\centerline {\epsfxsize=0.95\hsize \epsffile{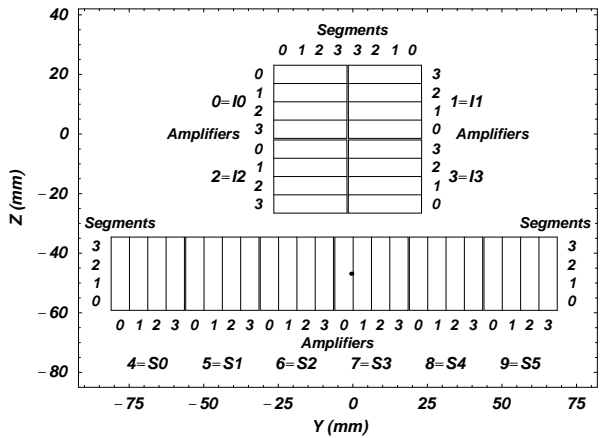} }
\caption {\label{fig:segments} 
ACIS detector layout showing CCD, amplifier, and segment identifiers.  The amplifiers are read out individually, and run horizontally for the I array CCDs and vertically for the S array CCDs.  For the spectral and FCC position registration analysis, each amplifier was divided into 4 segments.  Also shown by the black dot in amplifier 0 of CCD S3 is the position of the optical axis for the XRCF FCM measurement described in this paper.}
\end {figure}

Spectral analysis of FCM data taken at XRCF includes TRW IDs H-IAS-RC-1.001, H-IAS-RC-1.005, H-IAS-RC-1.004, and H-IAS-RC-1.002 for the I array electronic configuration, and H-IAS-RC--8.001 for the S array electronic configuration. 
RC-1.001 contains $\sim$12,900 s of data.  RC-1.004, RC-1.004, and RC-1.002 together contain $\sim$ 7,600 s of data and so were combined for all analyses discussed below.
RC-8.001 contains only $\sim$1,350 s of data, so the S array configuration data from XRCF are of very limited utility.
Spectral analysis of FCM data taken during OAC includes OBSIDs 62743 for the I array configuration and 62742 for the S array configuration.
Longer integrations were necessary to compensate for radioactive decay of the FCM sources, so 62743 contains $\sim$29,900 s of data and 62742 contains $\sim$ 18,300 s of data.
Regardless of the readout configuration, all data were taken with the optical axis at the nominal S array aim point (Fig.~\ref{fig:segments}), and with the ACIS detector at a position along the optical axis closest to the expected on-orbit position.  
The six readout CCDs for the I array configuration were all four I array front-illuminated CCDs and the two S array CCDs S2 (front-illuminated) and S3 (back-illuminated), and for the S array configuration all six S array CCDs (four front-illuminated and two, S1 and S3, back-illuminated).

\begin {figure} [htb]
\centerline {\epsfxsize=0.95\hsize \epsffile{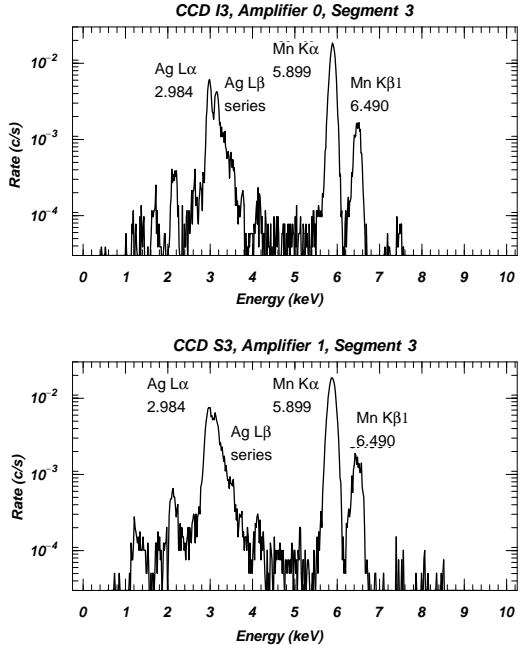} }
\caption {\label{fig:spectra} 
FCM/HRMA/ACIS spectra, from TRW ID H-IAS-RC-1.001, for a selected segment on the front-side-illuminated CCD I3 (top), and one on the back-side-illuminated CCD S3 (bottom).}
\end {figure}

ASCA grades 0, 2, 3, 4 and 6 were selected for further analysis.  
Using the line list~\cite{Bearden1967} given in Table~\ref{tab:lines}, fits with four free parameters plus the norm for each line determine the rates in each line for each CCD, amplifier and segment (Fig.~\ref{fig:segments}).
The 96 data points per line resulting from this procedure are sufficient for determining FCC position shifts.  
Figure \S\ref{fig:spectra} shows the fitted RC-1.001 spectra for a selected segment on the front-side-illuminated CCD I3 and one on the back-side-illuminated CCD S3.  
These segments lie near one another and above the optical axis (Fig.~\ref{fig:segments}).
The strongest lines in these spectra are the electron-capture lines Ag L$\alpha$, the Ag L$\beta$ series, Mn K$\alpha$, and Mn K$\beta$1.  
The superior energy resolution of the front-side-illuminated CCDs at the time of OAC is apparent. 
Comparison of XRCF line rates with those from OAC, corrected for radioactive decay, ACIS dead time and FCC position shifts, comprises the verification of the Chandra absolute flux calibration transfer and provides the basis for the search for any possible changes in molecular contamination.

\begin {table} [htb]
\caption {Line list for spectral fits.}
\label{tab:lines}
\begin {center}
\begin {tabular}{|c|c||c|c|} \hline
{\bf {Line}} & {\bf {Energy}} & {\bf {Line}} & {\bf {Energy}}\\ 
\rule[-1ex]{0pt}{3.5ex}   & \bf {(keV)} &   & \bf {(keV)} \\
\hline
\rule[-1ex]{0pt}{3.5ex}  (Ag L$\alpha$)${}^*$ & 1.2425 &  Ag L$\gamma$1  & 3.5226 \\
\hline
\rule[-1ex]{0pt}{3.5ex}  (Ag L$\beta$1)${}^*$ & 1.4111 & Ag L$\gamma$2 & 3.7432 \\
\hline 
\rule[-1ex]{0pt}{3.5ex}  Si K$\alpha$ & 1.7398 & (Mn K$\alpha$)${}^*$ & 4.1553 \\
\hline
\rule[-1ex]{0pt}{3.5ex}  Au M$\alpha$ & 2.1213 & Mn K$\alpha$ & 5.8951  \\
\hline
\rule[-1ex]{0pt}{3.5ex}  Au M$\beta$ & 2.2050 & Fe K$\alpha$ & 6.4000  \\
\hline
\rule[-1ex]{0pt}{3.5ex}  Ag L$\ell$ & 2.6637 & Mn K$\beta$1 & 6.4904  \\
\hline
\rule[-1ex]{0pt}{3.5ex}  Ag L$\eta$ & 2.8061 & Ni K$\alpha$ & 7.4724  \\
\hline
\rule[-1ex]{0pt}{3.5ex}  Ag L$\alpha$ & 2.9823 & Cu K$\alpha$ & 8.0411  \\
\hline
\rule[-1ex]{0pt}{3.5ex}  Ag L$\beta$1 & 3.1509 & Au L$\ell$ & 8.4939  \\
\hline
\rule[-1ex]{0pt}{3.5ex}  Ag L$\beta$6 & 3.2560 & Au L$\alpha$ & 9.7130  \\
\hline
\rule[-1ex]{0pt}{3.5ex}  Ag L$\beta$2 & 3.3478 & &  \\
\hline
\end {tabular}
\end {center}
\begin {itemize}
\item [${}^*$] Si K escape peak.
\end {itemize}
\end {table}

\section {FCM registration analysis} \label{s:regis}

In order to compare the FCM measurements at XRCF and OAC with each other and with ray trace simulations, the following differences between XRCF and OAC must be taken into account:

\begin{itemize}
\item [(1)]  The FCM/FCC/HRMA and ACIS orientation at XRCF differs by 180 degrees from that at OAC.
\item [(2)]  The gravity environment at XRCF (one g with compensation) differs from that at OAC (zero g without compensation) leading to small changes in figure.
\item [(3)]  The thermal environment at XRCF was $\sim$30 F warmer than at OAC leading to different thermal contraction of the FCC holding the FCM.
\item [(4)]  Since the FCC was removed after XRCF and then remounted, the relative position of the FCM/FCC and HRMA may have changed.  
During XRCF testing, the FCC V-groove-block and cup-cone positioning mechanism was not aligned, and the magnitude of the resulting displacement, while visually estimated at $\sim$0.5 mm mostly in Z, is not known accurately.  
Launch vibrations may also have contributed to changes in FCC positioning.
\item [(5)]  The OAC rates must be corrected for radioactive decay and ACIS detector dead time.
\end{itemize}

Items (1)-(4) require registration of the FCM images from XRCF and OAC by comparing measured images, corrected using the ACIS flat fields taken at XRCF, at chosen line energies with simulated images created using the 
Eastman-Kodak measured FCM source positions on the FCC with respect to its nominal axis~\cite{Schwab1996},
the as-built telescope model documented in the Chandra Mission Support Team calibration report,
and the ray trace code developed at MSFC~\cite{Elsner1998} with varying relative positions between the FCC and HRMA.  
We carry out the cross-correlation registration analysis at the strongest lines, Ag L$\alpha$ and Mn K$\alpha$ (E = 2.984 and 5.899 keV, respectively), letting the overall normalizations for each CCD assume values that minimize the value of $\chi^{2}$ at the various trial positions of the FCC.  
We also carry out the analysis at the rate weighted average energy of the Ag L line group (E $\simeq$ 3.198) and at Mn K$\beta$1 (E = 6.490 keV).

\begin {figure} [htb]
\centerline {\epsfxsize=0.95\hsize \epsffile{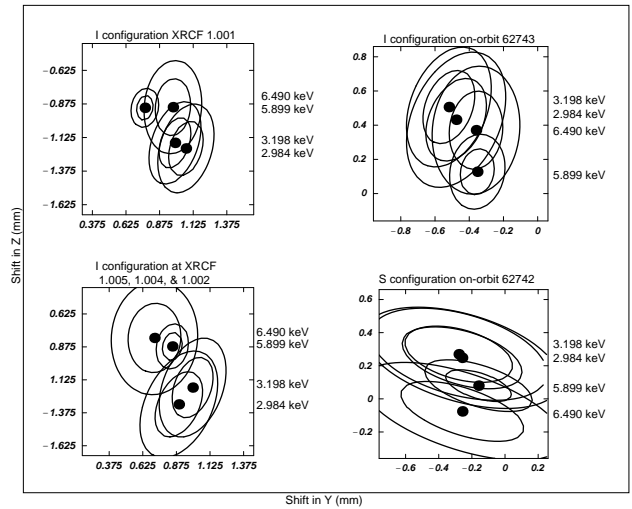} }
\caption {\label{fig:registration} 
Best fit values with error contours for the shift in FCC position from its nominal value relative to the HRMA.  The lines show the 67\% and 95\% two parameter error contours.}
\end {figure}

Although the ray-trace images generated using the Project Science ray trace code appear to adequately reproduce the FCM images, in fact the best-fits are statistically unacceptable due to a variety of systematic effects that are difficult to take fully into account.
Error contours were therefore constructed by rescaling $\chi^2$ values to 1 per degree of freedom at the best-fit~\cite{Press1992}.  
Fig.~\ref{fig:registration} shows the best fit FCC position shifts at Ag L$\alpha$, the Ag L$\beta$ group, Mn K$\alpha$ and Mn K$\beta$1, together with 67\% and 95\% two parameter error contours.  
The registration results for the XRCF and OAC cases show adequate agreement as a function of energy, and we now continue on to examine constraints on changes in the amount of molecular contamination on the HRMA from XRCF to OAC.

\section {Constraints on Molecular Contamination} \label{s:constraints}

After correcting for radioactive decay and the assumed ACIS on-orbit dead time of 2.5\%, and summing over all six CCDs, the percent change (positive meaning the OAC number is larger than the XRCF one) in total I array rate for OAC OBSID 62743 relative to the two XRCF rates is given in Table~\ref{tab:changes}.
The assigned errors are 1$\sigma$ statistical errors based on counts and do not include possible systematic contributions.

\begin {table} [htb]
\caption {Percent change in total I array FCM rate from XRCF to OAC (OBSID 62473 corrected for radioactive decay and ACIS dead time).}
\label{tab:changes}
\begin {center}
\begin {tabular}{|c|c|c|} \hline
{\bf }  Energy  & RC-1.001 & RC-(5,4,2).001 \\ 
\hline
\rule[-1ex]{0pt}{3.5ex}  2.984  & 1.80$\pm$0.39   & 1.59$\pm$0.49  \\
\hline
\rule[-1ex]{0pt}{3.5ex}  3.198  & 2.38$\pm$0.38   & 2.27$\pm$0.47  \\
\hline
\rule[-1ex]{0pt}{3.5ex}  5.899  & -0.13$\pm$0.08  & 0.25$\pm$0.11  \\
\hline
\rule[-1ex]{0pt}{3.5ex}  6.490  & 0.19$\pm$0.93   & 0.16$\pm$1.25  \\
\hline
\end {tabular}
\end {center}
\end {table}

\begin {figure} [htb]
\centerline {\epsfxsize=0.95\hsize \epsffile{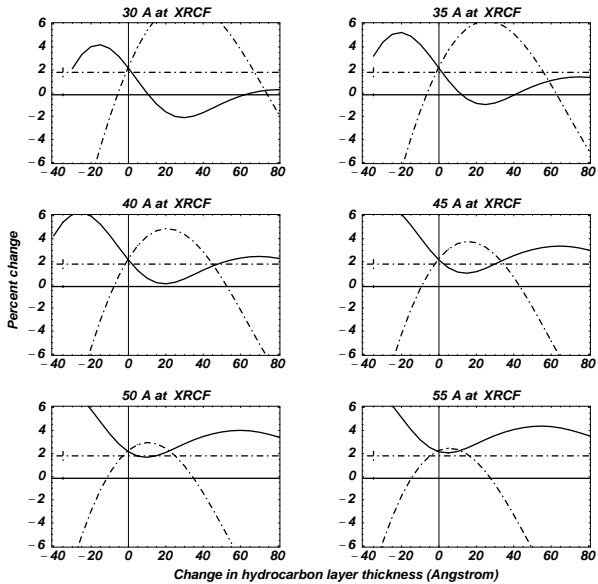} }
\caption {\label{fig:changes} 
Measured (horizontal lines) and predicted percent changes in total ACIS I array configuration rate from XRCF to OAC, at 2.984 (dash-dotted lines) and 5.899 (solid lines) keV, as a function of assumed change in the thickness of a putative hydrocarbon layer. 
The different panels are for different assumed amounts of hydrocarbon present at XRCF.}
\end {figure}

Predicted changes in FCM/HRMA/ACIS rate due to changes in molecular contamination from XRCF to OAC depend on the amount assumed to be present at XRCF, as shown in Fig.~\ref{fig:changes}.
The composition of the hydrocarbon layer assumed for these simulations was four hydrogen atoms for every carbon atom with a bulk density of 1 g/cm$^{3}$.
The overall shape of the these predicted curves remains approximately constant but their relative positioning depends dramatically on the amount of hydrocarbon assumed present at XRCF.

Our analysis is much refined from that reported two years ago~\cite{Elsner1998}, but further work is required in the following areas:

\begin{itemize}
\item [(1)]  We will perform a more thorough analysis of on-orbit ACIS dead time.
\item [(2)]  We will attempt the difficult task of achieving a thorough understanding of sources of systematic errors.
\end{itemize}

At the present time, we believe our results support the following statements:

\begin{itemize}
\item [(1)]  The predicted maximum change in molecular contamination from XRCF to OAC was 10 $\AA$~\cite{Ryan1999}.  Because of possible systmatic effects, we can not yet determine whether the actual change satisfies this limit.  
\item [(2)]  However, any deviations from the predicted on-orbit HRMA throughput due to changes in molecular contamination from that present at XRCF are less than a few percent even at the energies most sensitive to such changes.
\item [(3)]  Only a small amount of molecular contamination is present on the HRMA, both on-orbit and at XRCF.
\end{itemize}

\section {Acknowledgements} \label{s:acknowledge}

We wish to thank Terry Gaetz, Ping Zhao and Leon Van Speybroeck, and the SAO Chandra Mission Support Team as a whole, for their considerable help in understanding the as-built model for the HRMA.  Needless to say the successful construction, launch and operation of the Chandra X-ray Observatory resulted from the efforts of a very great number of dedicated people.  We thank them all.

\bibliography{../axaf}
\bibliographystyle{spiebib}

\end{document}